%% file: paperG597.tex
\documentclass[12pt,preprint]{aastex}
\usepackage{t1enc}
\usepackage{graphicx}
\usepackage{natbib}
\usepackage{lscape}
\newcommand{\hii}{\mbox{\ion{H}{2}}}

\newcommand{\kms}{km~s$^{-1}$}

\newcommand{\msun}{$M_{\odot}$}

\newcommand{\cmt}{cm$^{-3}$}

\newcommand{\jpb}   {$\rm Jy~beam^{-1}$}

\newcommand{\gap}%
{\raisebox{-0.5ex}{$\stackrel{\scriptstyle >}{\scriptstyle \sim}$}}

\shorttitle{VLA observations of M8}
\begin{document}








\title{Very Large Array and Jansky Very Large Array observations of the compact radio sources in M8}

\author{
Josep M. Masqu\'e\altaffilmark{1},
Sergio Dzib\altaffilmark{2}, \&
Luis F. Rodr\'iguez\altaffilmark{1,3}
} 

\altaffiltext{1}{Centro de Radioastronom\'ia y Astrof\'isica, Universidad Nacional Aut\'onoma de M\'exico, Morelia 58089, M\'exico}

\altaffiltext{2}{Max-Plank-Institut f\"{u}r Radioastronomie, Auf dem H\"{u}gel 69, D-53121 Bonn, Germany}

\altaffiltext{3}{Astronomy Departament, Faculty of Science, King Abdulaziz University, P.O. Box 80203, Jeddah 21589, Saudi Arabia}

\begin{abstract}

We analyze high-resolution Very Large Array and Jansky Very Large Array continuum observations of the M8 region carried out at several epochs that span a period of 30~yr. Our maps reveal two compact sources. One is associated with Her~36 SE, a possible companion of the O7 luminous massive star Her~36, and the other is associated to G5.97-1.17, whose proplyd nature was previously established. With the analyzed data, we do not find significant time variability in any of these sources. The derived spectral index of $\ge 0.1$ for Her~36 SE, the marginal offset of the radio emission with the previous IR detection and the associated X-ray emission previously reported suggest the presence of an unresolved interaction region between the strong winds of Her~36 and Her~36 SE. This region would produce non-thermal contamination to the global wind emission of Her~36 flattening its spectral index. On the other hand, the emission of G5.97-1.17 can be  also explained by a mixture of thermal and non-thermal emission components, with different relative contribution of both emission mechanisms along the proplyd. We argue that the shock created by the photo-evaporation flow of the proplyd with the collimated stellar wind of Her~36 accelerates charged particles in G5.97-1.17 producing considerable synchrotron emission. On the contrary, an electron density enhancement at the southwest of G5.97-1.17 makes the thermal emission dominant over this region.  

\end{abstract}

\section{Introduction}

The intense radiation produced by recently formed massive stars is capable of ionizing their surrounding material, giving rise to large (> 0.1~pc) photo-ionized regions (\hii\ regions). In the \hii\ regions, the free electrons emit thermal radiation through the free-free mechanism, which produces bright emission at radio-wavelengths being easily detected with traditional interferometers. As a result, large surveys of massive star-forming regions in the Galaxy can be efficiently performed at cm wavelengths and the properties of \hii\ regions, as well as of their compact precursors called ultra-compact and hyper-compact \hii\ regions (UC\hii\ and HC\hii, respectively), are nowadays well established \citep{wood1989, becker1994, hoare2012}.

Recently, special attention has been called to extremely compact objects projected inside UC\hii\ regions hidden by the bright extended cm emission of the ionized gas \citep[e.g. W3(OH), NGC6334A:][]{kawamura1998, carral2002,dzib2013a,rodriguez2014,dzib2014}. The behavior of these compact sources is puzzling: their time variability in scales of months to years and large fluxes are discrepant with those expected for a free-free emitting stellar wind. Moreover, it seems to be no unique explanation for their nature: while the compact source in the center of W3(OH) exhibits a positive spectral index consistent with the presence of a fossil photo-evaporating disk \citep{dzib2013a}, the compact source of NGC6334A has a negative spectral index characteristic of optically thin synchrotron emission \citep{rodriguez2014}. The last emission mechanism requires the presence of relativistic electrons, which can be attributed to Fermi acceleration in strong shocks of stellar winds between O companions, implying the presence of a massive binary system ionizing the HII region \citep[e.g. several sources in the Cyg OB2 association:][]{ortizleon2011,blomme2013}. 

A technique employed to detect these compact sources consists in discarding the visibilities from the shortest baselines when making the map. These baselines contain information of the extended emission and, hence, the resulting map obtained under this technique retains only emission from the most compact objects \citep[see][for an example]{dzib2013a}. This technique has been applied successfully for the UC~\hii\ region W3(OH) \citep{dzib2013a} and NGC6336A \citep{carral2002}. In this paper we apply the same technique to the NGC 6523 massive star forming region.

NGC 6523 is an extremely dense stellar cluster belonging to the OB1 Sagittarius association. Observations to the date reveal a complex scenario for this region: it appears to be located at the front edge of a molecular cloud, whose foreground material has been blown up by the stellar activity of NGC 6523 and a new generation of stars is being triggered inward the cloud \citep{lada1976, elmegreen1977, lightfoot1984}.
Behind this cluster, projected in the center, we find the famous Lagoon Nebula, or M8, an HII region, located at 1.3~kpc \citep{arias2006}, that is sustained by the UV photons from the O7 star Herschel 36 \citep[][hereafter Her~36]{woolf1961}. This luminous star, detected previously at optical and IR wavelengths, has multiple components \citep{allen1986,arias2010}. 
The gas ionized by Her~36 forms a blister \hii\ region shaped by the complex distribution of material in the region (but see \citealt{lightfoot1984} for caveats to this interpretation), acquiring a distinctive morphology which lead to call it as the 'Hourglass Nebula' \citep{allen1986}. Radio observations presented by \citet{stecklum1998} show diffuse emission spanning over a large area, but no significant emission coming directly from Her~36. These observations also show that the brightest peak of the diffuse radio emission falls 2\rlap{$''$}.7 to the southeast of the massive star. This bright emission was first associated by \citet{wood1989} to an UCHII region ionized by a B0 star (it was called G5.97$-$1.17, and we will refer to it as G5.97 through the rest of the paper). However, later studies showed that G5.97 is most likely a proplyd being photoevaporated by Her~36 \citep{stecklum1998}. Up to date, the majority of currently known proplyds are located in the Orion nebula \citep{ricci2008}, which makes the few proplyd candidates in other \hii\ regions a good case of study, since their potential proplyd nature may have different intrinsic properties and associated ambient. Indeed, the technique mentioned above is ideal to isolate the compact cm emission of proplyds since they are generally embedded in an extended ionized strong emitting ambient.

In this paper we analyze archival Very Large Array (VLA) and new Jansky Very Large Array (JVLA) data of continuum cm emission observed toward the M8 region. We were aimed to determine the nature of the compact sources found in the region. In Section 2 we describe the archival observations. In Section 3 we present the resulting maps and derive spectral indices. In Section 4 we discuss a possible interpretation of the physical processes involved in the detected sources. Finally, in Section 5 we give a summary of our conclusions. 

\section{Observations}

The separation between Her~36 and the G5.97 source in the plane of sky is 2\rlap{$''$}.7. Thus, high angular resolution observations carried out in the A or B configuration are required, which have enough angular resolution to separate both sources. On the other hand, the fluxes of the compact sources projected inside UCHII regions reported to date are typically of the order of a few mJy at cm wavelengths \citep{carral2002,dzib2013a,rodriguez2014}. Therefore, the second criterion is to select deep observations providing maps with an $rms$ noise below 1~mJy. We inspected the VLA archive looking for observations accomplishing both conditions and found a handful of projects whose observing epochs span nearly 20 yr. These data were calibrated using the Common Astronomy Software Applications (CASA), except for the January 1988 and December 1996 data, that were calibrated with the Astronomical Image Processing System (AIPS).  

Additionally, recent observations of M8 were carried out with the Jansky Very Large Array in the A configuration at K band on 2014 May, as part of a project aimed to look for compact sources projected inside UCHII regions. We used the dual-polarization mode that provides 64 spectral windows of 128~MHz giving a total bandwidth of 8~GHz, much wider than the archival observations presented above (typically 100~MHz). These data were calibrated using the CASA package. A summary of the parameters of the VLA and JVLA observations is given in Table~\ref{observations}.

\section{Results}

\subsection{Emission Maps}

\begin{figure}[htbp]
\centering
\resizebox{1.5\textwidth}{!}{\includegraphics[angle=0.]{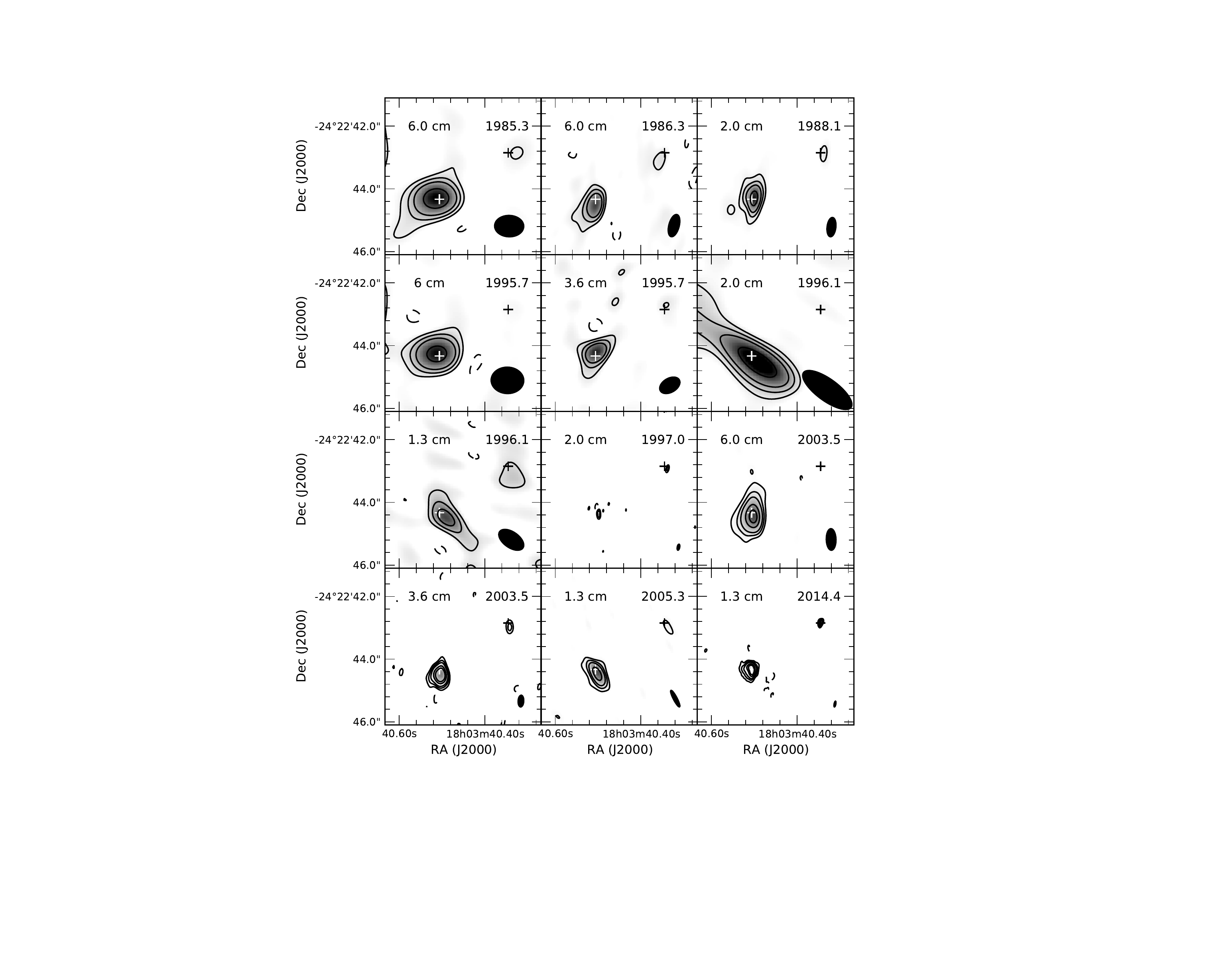}}
\vspace{-2cm}
\caption{
VLA emission maps of the G5.97 region observed at several bands and epochs (gray scale and contours). Contours are -4,-3, 3, 4, 5, 7, 10, 15 and 20 times the $rms$ noise level of each map (see Table~\ref{observations}). The beam is shown in the bottom right corner. The two crosses show the position of G5.97 (SE) and Her~36 (NW) derived in the map obtained with superuniform weighting of the observations carried out at 1.3~cm on May 2014 shown in Fig.~\ref{superuniform}.\label{G597total}}
\end{figure}

The corresponding natural-weighted maps to the observations of the G5.97 region listed in Table~\ref{observations} are shown in Figure~\ref{G597total}. The maps were constructed discarding the short spacings of the uv-dataset in order to remove the large scale emission. In particular we removed the baselines shorter than 65~k$\lambda$ at 6~cm, 100 k$\lambda$ at 3.6~cm, 150~k$\lambda$ at 2~cm and 90~k$\lambda$ at 1.3~cm, which corresponds to angular scales of 3\rlap{$''$}.2, 2\rlap{$''$}.1, 1\rlap{$''$}.4 and 2\rlap{$''$}.1, respectively. These cuts of the uvrange were chosen based on a trade off between removing extended emission and keeping enough visibilities to obtain a good map.
The remaining emission shows two compact radio-sources, one to the southeast and another to the northwest of the mapped region, corresponding to G5.97 and Her~36, respectively. 

Emission associated with Her~36 is clearly detected as a compact radio source in these images. It appears always as a point source at 1.3 cm in May 2005 and May 2014, at 2~cm in January 1988 and December 1996, and in the 3.6~cm observations of July 2003. On the other hand, it is marginally detected at 6~cm in the observations of April 1985 and 1986, and undetected in the rest of the observations. The maps of September 1995 and January 1996 have less quality with respect to the other maps and are not appropriate for the Her~36 detection. Also, the poor detectability of Her~36 when observed at 6~cm compared to shorter wavelengths suggests a positive spectral index for this source. We will discuss further the spectral index of Her~36 in the next section. As we will see later, this radio emission is not directly associated with Her 36, but is located between this star and a very nearby companion.

The G5.97 source is detected in all the bands and all the epochs, it also appears to be partially resolved. At 6 cm, the proplyd is marginally elongated towards the southeast, opposite to Her~36, specially in the 1985 observations where the elongation is seen as a tail. At 3.6 cm, G5.97 appears to be composed of several blended components. This is best seen in the July 2003 map, which has the highest angular resolution of the 3.6 band, where 
the source presents two components: one component can be associated with the southeastern elongation discussed above and the other component is elongated roughly in the north-south direction. At 2 cm and 1.3 cm, G5.97 appears more compact compared with the maps corresponding to longer wavelengths and the southeastern elongation is hardly noticed.
In particular, the May 2014 observations at 1.3~cm, carried out with very high angular resolution capable of resolving G5.97, reveal a cometary morphology with the tip of this shape pointing to Her~36. 

\subsection{Fluxes and Spectral Indices}

In Table~\ref{fluxHer36} we report the flux peak value and position of the radio source associated with Her~36 . We give as upper limits of the peak flux for the non-detections and marginal detections 3 times the $rms$ noise of the maps. In Table~\ref{fluxG597} we present the parameters of the emission of G5.97 derived in the following manner: we first used the outcome of the 'findsources' function of CASA to set a reliable fitting region around each source; then we used the 'fitcomponents' function in each region to fit a Gaussian plus a zero-level offset and extract the parameters shown in the Table. A close inspection of the flux values of the Tables shows no or modest time variability for both sources (less than 30\%). Since we are comparing data taken at different epochs with different observational setups (e.g., array configuration, uv-coverage, calibration, etc.), we consider as not significant any time variability below 30\%. For the same reason, the estimation of spectral indices is only reliable with data corresponding to the September 1995, January 1996 and July 2003 observations, when two bands were simultaneously observed. In these epochs, Her~36 is only detected at 3.6~cm in July 2003 allowing us to derive a lower limit of the spectral index for this epoch ($\geq 0.1$). The poor detection of Her~36 at 6~cm favors a dominant thermal nature for the Her~36 emission. On the other hand, G5.97 has a flat spectral index: by using the 3.6 and 6~cm bands we derived a spectral index of $-0.05 \pm 0.19$ in September 1995 and $0.02 \pm 0.28$ in July 2003. The 1.3 and 2.0~cm data of the observations of January 1996 yield $-0.14 \pm 0.55$, which is more uncertain due to the lower quality of the maps of this epoch but still consistent with a flat spectral index.

In order to study the spectral index derived as a function of position, we obtained the map of the spectral index of the July 2003 observations combining the emission maps at 6 and 3.6 cm, which is shown in Figure~\ref{spectral_index}. We did not show the spectral index maps of the January 1996 and September 1995 epochs, where also several bands where simultaneously observed, because the poor quality of the emission maps prevented us to obtain any significant trend. The initial maps were constructed with the same UV-range and restoring beam (corresponding to the largest beam of the two combined maps, i.e. the map at 6~cm), and a cutoff at 7 times the $rms$ noise level. This cutoff is adequate to derive a significant trend for the spectral index in the map and only derives the uncertain extreme negative value of -1 at the north of G5.97. The most noticeable feature is the gradient of spectral index over the NE-SW direction. This gradient can be explained if we consider the cm emission of G5.97 as a mixture of thermal or non-thermal emission: there is different contribution of each type of emission in different parts of G5.97.

\section{Discussion}

\subsection{The Compact Radio Emission of Her~36}

The positive spectral index of Her~36 derived above ($\geq0.1$) suggests that we are detecting the radio emission of a free-free ionized stellar wind with an optically thick portion. However, if we assume that the wind is emanating from an O7 V star, a flux of 0.026~mJy is expected at 3.6~cm according to the tabulated values of \citet{dzib2013b} scaled at 1.3~kpc of distance, which is clearly below the flux measured for Her~36, when detected at 3.6 cm (0.91~mJy). Following the formulation of \citet{panagia1975} and adopting a terminal velocity for the stellar wind of 2700~\kms  \citep{dzib2013b}, the fluxes of Her~36 at 3.6, 2 and 1.3 cm yield to values between $(1.5-2.1) \times 10^{-5}$~\msun~yr$^{-1} $ for the mass loss rate of the massive protostar, more than one order of magnitude above the expected values derived from \citet{vink2001} assuming that Her~36 has a solar metallicity. 

High resolution infrared images of Her~36 \citep{goto2006,stecklum1995} reveal an additional compact source, located 0\rlap{$''$}.25 SE (hereafter Her~36 SE) of the Her~36 luminous star. \citet{goto2006} detected this compact source also at 2~cm and attributed this emission to an embedded early B type star that produces a small HII region,  even though they also discussed the influence of the nearby Her~36 luminous star in terms, for instance, of external photoionization (i.e. a proplyd or photoevaporating globule). In the top panel of Figure~\ref{superuniform} we show the 1.3~cm emission around Her~36 obtained with superuniform weighting, which appears to be associated rather with the Her~36 SE component, consistent with the \citet{goto2006} results. Thus, the discrepancy of the radio flux derived above is explained if we assume that it comes from a process occurring near Her~36 SE instead of being associated to the wind directly emanating from the Her~36 main component. Our derived lower limit for the spectral index ($\geq 0.1$) is consistent either with the presence of a small HII region (the expected spectral indices can be flat or positive up to 2) and a proplyd or photoevaporating globule (with an expected thermal flat spectral index, e.g. see \citealt{felli1993}) in Her~36 SE. The required Lyman photons to have a flux of 1.22~mJy at 1.3~cm for Her~36 SE is $2.1 \times 10^{44}$~ photons~s$^{-1}$, easily provided by an early B star embedded in Her~36 SE, or an O7 star separated 325~AU (assuming 1.3~kpc of distance and 0\rlap{$''$}.25 for the size of Her~36 SE that gives a geometric dilution factor of 0.06). Therefore, with the present data we are not able to discard neither internal nor external ionization for Her~36 SE. 

However, a close inspection of the high resolution map of Her~36 of Fig.~\ref{superuniform} shows that, within the errors, the 1.3~cm emission is slightly offset towards the direction of the Her~36 main component. If this offset is significant, the radio emission of Her~36 SE could be ascribed to a wind interaction region, presumably fairly close to Her~36 SE since its wind is expected to be significantly less powerful than that of the Her~36 main component. The lower limit for the spectral index of Her~36 is compatible with a flat index expected from a mixture of free-free emission of the ionized material of the wind and a non-thermal component produced in the wind collision region. A similar scenario was found in the massive multiple system Cyg OB2 \#5 of the Cygnus OB2 association \citep{contreras1997,kennedy2010,ortizleon2011,dzib2013b}. However, the emission arising from the wind collision region associated to Cyg OB2 \#5 was found to be variable in a period of a few years. This variability was attributed to variations of free-free opacity as the companion moves along the orbit and the column density of ionized material changes with the line of sight. If we consider that Her~36 SE is orbiting around the Her~36 main component, adopting their separation as the orbital radius and 30~\msun\ for Her~36, we get a period of 1200~yr. Therefore, in the case 
that the radio emission arises from a wind colliding region, in 30~yr we do not expect significant changes in the observed flux of Her~36. 

Besides, \citet{rauw2002} reported the detection of X-ray emission peaking at Her~36 but also presenting a distant enhancement located $\sim0.2$~pc far away at the southeast of Her~36. 
The presence of X-ray emission in Her~36 is consistent with the existence of a wind interaction region between the Her~36 main component and Her~36 SE. Also, these winds must be powerful enough to interact with the hosting molecular cloud, $\sim0.2$~pc far away from Her~36, producing hot gas and creating the extended southeastern enhancement of X-ray emission \citep{rauw2002}. Very high resolution (e.g., VLBI) observations are required to resolve and confirm the existence of a wind colliding region in Her~36, and assess if the involved winds are powerful enough to alter the hosting cloud and other associated objects, such as the G5.97 proplyd.


\subsection{Thermal and Non-Thermal Emission in G5.97}

In a previous study \citet{stecklum1998} provide evidence for the proplyd nature of G5.97.
The fluxes presented in Table~\ref{fluxHer36} and \ref{fluxG597} are consistent with the proplyd hypothesis. First, the flux values range between 13 and 19~mJy in all the observed bands and epochs, implying a rather flat average spectral index, and consistent with the 17~mJy measured by \citet{stecklum1998} at 2~cm. These authors estimated the number of required Lyman photons to provide this flux and found a slight deficit of ionizing photons coming from Her~36, even though the application of the main-sequence of Lyman luminosity to the young Her~36 star could not be appropriate at all. Another property supporting the proplyd hypothesis is the measured steady flux of G5.97. The timescale for dissipation of a proplyd is of the order of $10^4$ years \citep{henney1999,miotello2012} and we do not expect significant changes in a few years. Besides, \citet{castelaz1995} compared HST images of the Orion proplyds,
finding no evidence of morphological or flux variations. In their VLA study of Orion at 3.6 cm, \citet{zapata2004} detected a total of 77 compact radio continuum sources, of which 30 are associated
with proplyds. While most of the sources associated with proplyds appear to be
steady in time, nine of them exhibit significant flux density
variations along the 4 years of the Zapata et al. study.
Does this imply that the free-free emission from the proplyds is
variable in time? As noted by \citet{zapata2004} the radio continuum emission
from a proplyd may be contaminated by time-variable non-thermal emission
from the associated young star, as indicated by the evidence of circular polarization found at least for one source (their source number 6). We mapped the Stokes V parameter to look for circular polarization in all the observations of Table~\ref{observations} to search for gyrosynchrotron emission in G5.97 with null detections. Possibly, at the distance of M8, the intensity of the circular polarized emission, if present, is expected to be extremely weak. Thus, this non-detection is consistent with the fact that G5.97 does not have a variable non-thermal component, i.e., coronal emission.



\begin{figure}[htbp]
\centering
\resizebox{1.2\textwidth}{!}{\includegraphics[angle=0.]{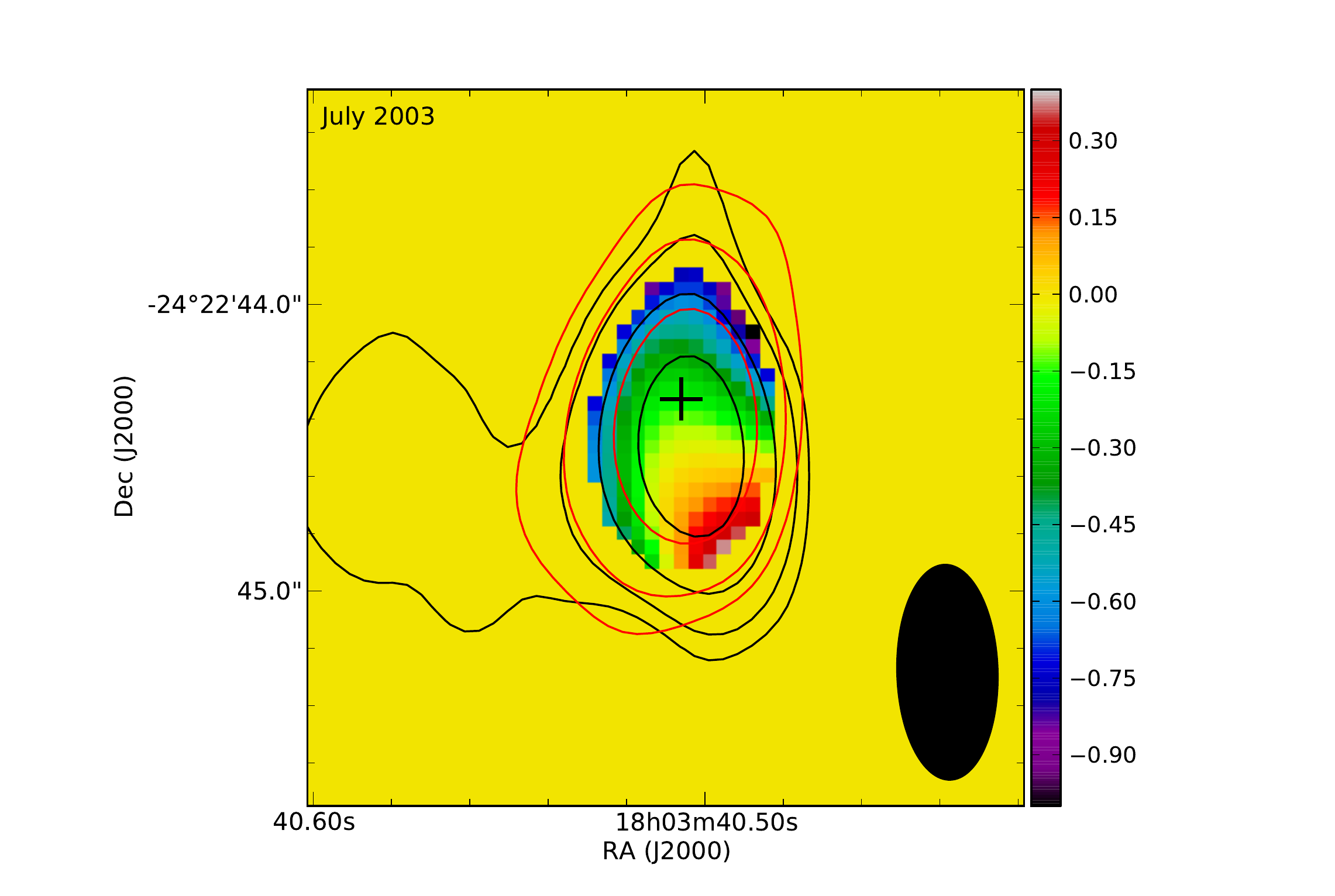}}
\caption{
Maps of spectral index distribution in G5.97 obtained with 
the 6 (red contours) and 3.6 cm (black contours) bands observed during July 2003. The maps were obtained with the same UV range (from 50 to 600 k$\lambda$) and restored with the same synthesized beam ($0.75\arcsec \times 0.35\arcsec$ and P.A. 1.5\arcdeg). The contours 
are always 3, 5, 10 and 20 times the $rms$ noise of each map (0.32 mJy 3.6 cm and 0.65 mJy at 6 cm). The cross marks the G5.97 position shown in Fig.~\ref{G597total}. The beam is shown in the bottom right corner. \label{spectral_index}}
\end{figure}

\begin{figure}[htbp]
\centering
\resizebox{0.6\textwidth}{!}{\includegraphics[angle=0.]{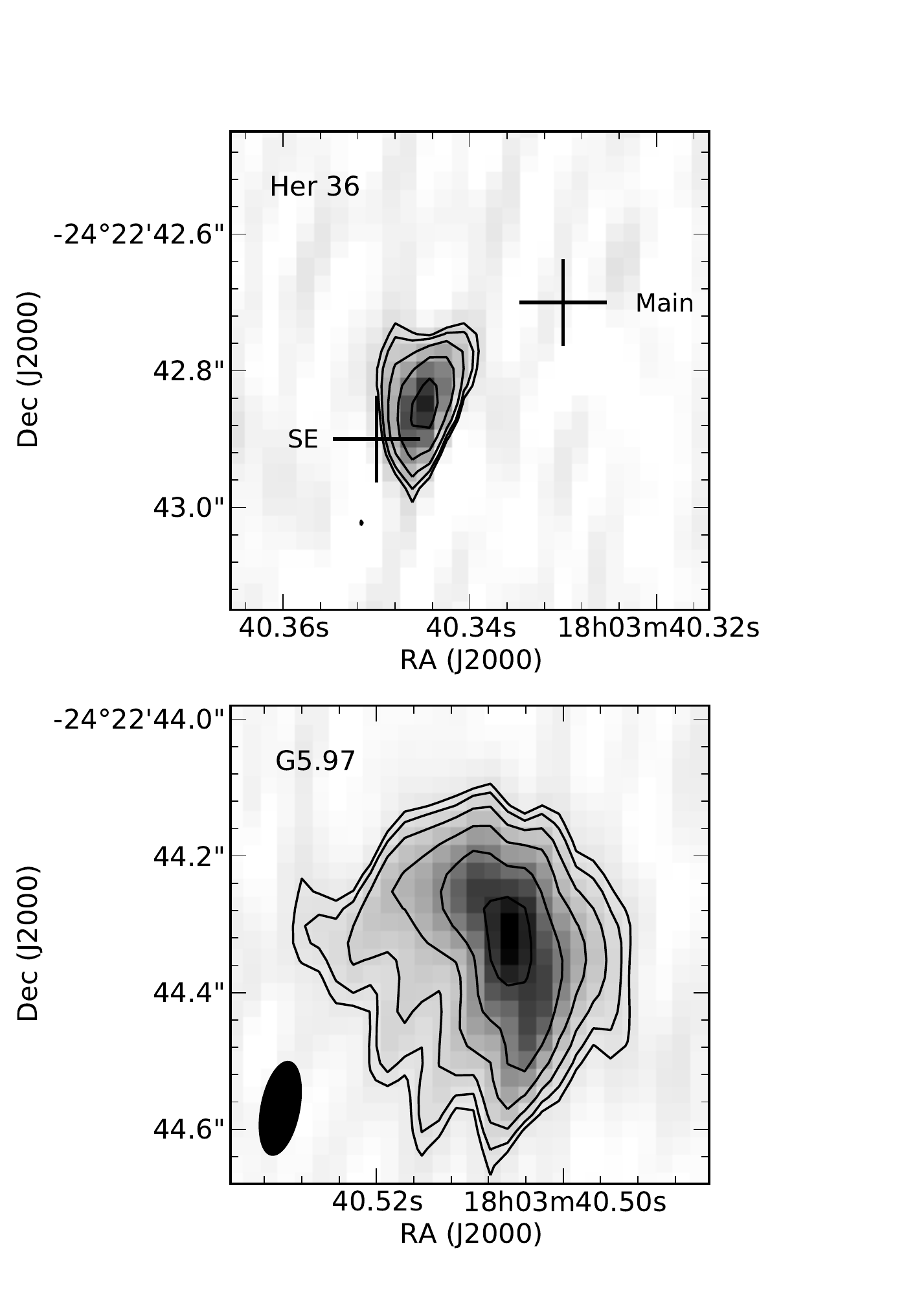}}
\caption{
\emph{Upper panel}: Zoom around Her~36 of the 1.3~cm continuum map corresponding to the observations of 2014 May constructed with superuniform weighting (gray scale and contours). The two crosses corresponds to the SE (southeast) and Main (northwest) components of Her 36. \emph{Lower panel}: The same as the upper panel but around the G5.97 proplyd. In both panels, contours are -3, 3, 4, 6, 10, 18 and 30 times the $rms$ noise level of the map (0.45~m\jpb). The synthesized beam of the map is shown in the bottom left corner of the lower panel. \label{superuniform}}
\end{figure}

Despite its assumed proplyd nature, G5.97 exhibits some differences with the 'classical' proplyds found in the Orion Nebula. In addition to having a higher mass loss rate and being about 3 times larger than the largest Orion proplyds \citep{stecklum1998}, G5.97 may experience different physical processes. Figure~\ref{spectral_index} shows that the average flat spectral index of G5.97, a typical feature of most proplyds, is a combination of thermal and non-thermal emission. The fact that in the figure the peak in the spectral index is displaced with respect to the emission peaks (and these are also displaced between them), can be explained with a scenario where there is a region of thermal emission surrounded by extended emission with an important non-thermal contribution. Thus, the spatial distribution of the cm emission in G5.97 is the result of several components. Similar results were found previously for proplyd-like objects in NGC 3606 \citep{mucke2002}. For this case, synchrotron emission from a relativistic particle population embedded in a magnetic field was proposed to explain the nature of the non-thermal emission. Among several possibilities, they argued that shocks created by the photo-evaporated flow of the proplyd could provide the required particle acceleration. In any case, the fact that a fraction of the measured flux of G5.97 corresponds to non-thermal emission provides another possible explanation for the discrepancy, found by \citet{stecklum1998}, between the required and received ionizing photons from Her~36.

Similarly to the proplyd-like objects of NGC~3606, the non-thermal extended emission in a considerable fraction of G5.97 can be explained by the presence of relativistic electrons in a magnetic field. The relativistic electrons can be accelerated by the shock created by the photo-evaporated flow of G5.97 with the strong collimated wind of Her~36 (see previous section). Alternatively, a stellar origin for the non-thermal emission seems unlikely, given the fact that it is extended in a big portion of G5.97, consistent with the non-detection of circular polarized emission. As in \citet{mucke2002}, the results presented in this paper prove that the magnetic field likely plays a role in the physical processes occurring in proplyds. Also, not only the UV photons but the wind coming from the ionizing star can also be responsible for, at least, part of the proplyd radio emission.

The thermal emission found at the SW of G5.97 has spectral index values up to 0.6, typical of thermal radio-jets. Thus, we first consider the presence of a possible protostellar jet powered by the star associated with G5.97 as the possible origin for the thermal emission. Jet signatures have been detected in proplyds with outflow velocities of 100~\kms\ \citep{henney2002}, being commonly less powerful compared to the jets powered by other young stellar objects \citep[e.g., see][]{anglada1996a}. Following the \citet{curiel1987,curiel1989} formulation for a shock wave caused by the jet impinging in the circumstellar material, under the Rayleigh-Jeans approximation, the total expected flux in milliJanskys is:

\begin{equation}
S_\nu=1.42 \times 10^2\left(\frac{\Omega}{\mathrm{arcsec}^2}\right)\left(\frac{\nu}{5~\mathrm{GHz}}\right)^2 \left(\frac{T_e}{10^4~\mathrm{K}}\right)[1-\mathrm{exp}\left(-\tau_{\nu}\right)], 
\end{equation} 

\noindent
where $\Omega$ is the angular size of the emitting area (i.e., the region affected by the jet), $T_e$ is the electronic temperature, $\nu$ is the frequency and $\tau_{\nu}$ is the optical depth given by the expression:

\begin{equation}
\tau_{\nu}=1.55 \times 10^{-3}\left(\frac{n_0}{10^4~\mathrm{cm}^{-3}}\right)\left(\frac{V}{100~\mathrm{km~s}^{-1}}\right)^{1.68}\left(\frac{\nu}{5~\mathrm{GHz}}\right)^{-2.1}\left(\frac{T_e}{10^4~\mathrm{K}}\right)^{-0.55},
\end{equation}

\noindent
where, $n_0$ and V are the pre-shock density and the velocity in which the jet impinges on the medium, respectively. Assuming that all the gas is ionized with $T_e=10^4$~K and adopting a pre-shock density of $(4-20) \times 10^4$~\cmt\ \citep{stecklum1998} and a typical jet velocity of 100~\kms\ \citep[e.g.,][]{henney2002}, we derive values of $\tau_{\nu}$ between 0.005 and 0.03 for the 6~cm band. For an area of 0.3~arcsec$^2$, which corresponds to the size of the restoring beam of the maps of Fig.~\ref{spectral_index}, 
equation (1) yields to predicted 6~cm flux densities of 0.2-1~mJy. On the other hand, integrating in the 6~cm map of Fig.~\ref{spectral_index} the flux arising from the same beamsize area used above (0.3~arcsec$^2$), which encloses the region where the spectral index is $> -0.1$ (i.e., the region of G5.97 were the thermal emission is dominant) and limited by the contour 3 times the $rms$ noise level at the southwest, we obtain 5~mJy. Applying the same calculations and assumptions at the 3.6~cm band we obtain the same flux discrepancy (0.3-1.7~mJy vs. 5~mJy measured in the 3.6~cm map). Thus, unless there is a jet in G5.97 with significantly higher mass loss rate and/or terminal velocity than the typical jets associated with proplyds and/or other assumptions made here are wrong, the thermal emission of G5.97 must have another origin. 
 
Other possibility is that the electronic density, $N_e$, increases at the SW of G5.97. Since thermal free-free emission is proportional to $N_e^2$ and synchrotron emission is proportional to $N_e$, an enhancement of the electron density high enough (and the gas density in the case of being fully ionized) would make the thermal contribution dominant. In this sense, \citet{mucke2002} argue that the higher electron densities found in the Orion proplyds with respect to the NGC 3606 proplyd-like objects makes the contribution of the thermal emission more important for the former. This is not the case of G5.97: in the lower panel of Fig.~\ref{superuniform}, G5.97 appears resolved and presents a bow-shock morphology, with the tip pointing to Her~36, located in a limiting region of the proplyd where the non-thermal emission starts to be dominant (spectral index $\lesssim 0.1$, see the peak position in Fig.~\ref{spectral_index}). This suggest that an important part of the radio emission arises with the interaction of G5.97 with the stellar wind coming from Her~36. This is also seen in the Figure 2a of \citet{stecklum1998}, obtained from the subtraction of a scaled H$\alpha$ image from the broadband $HST$ image of M8. Here, G5.97 shows an ionized bow oriented towards Her~36 and possibly associated with the bow-shock identified in Fig.~\ref{superuniform}, clearly separated from a star located behind, which is possibly associated with the disk of the proplyd. According to the orientation of G5.97 with respect to Her~36, the position of the disk plus associated star, where the largest concentration of material is probably found, are approximately coincident with the part of G5.97 with positive spectral index, while the rim and the rest of the source appear associated with the non-thermal emission. Although the picture proposed here implies a highly asymmetrical appearance of G5.97, some inclination angles of the circumstellar disk with respect to the photo-ionizing star can develop complex morphologies for the proplyds \citep{henney1996}. High resolution observations at mm wavelengths to map the mass distribution of G5.97 and, if possible, resolve the disk, are required to test the hypotheses discussed here.



\section{Summary and Conclusions}

In this paper we monitored the radio emission of M8 by analyzing archival VLA and new JVLA observations spanning about 30 yr at 1.3, 2, 3.6, and 6 cm bands. We detected the compact source Her~36 SE, a possible companion of the Her~36 luminous star. The emission is not resolved and appears slightly displaced towards the Her~36 main component. The derived spectral index of $\ge 0.1$ for the Her~36 star, being preferably detected at short wavelengths, is consistent with that of thermal emission with some non-thermal contamination. A possible explanation relies on  
the presence of an unresolved interaction region between winds of Her~36 and Her~36 SE, that would produce significant non-thermal emission, consistent with the previous detection of X-ray emission \citep{rauw2002} over this source. These winds are likely powerful enough to affect the surrounding cloud, including the G5.97 proplyd. 

The G5.97 source presents features consistent with having a proplyd nature. It is partially resolved, specially at long wavelengths bands, where a tail oriented in opposition to Her~36 is marginally seen at 6~cm, and with our highest resolution map at 1.3~cm, where a cometary  morphology pointing to Her~36 is observed. Its flux does not vary significantly with time with values ranging between 12.5 and 19~mJy in all the observed bands and epochs, also implying a rather flat spectral index. We did not detect circular polarized emission in G5.97 indicating that the contamination of gyrosynchrotron emission from the associated star is not important, consistent with the non-variability of the G5.97 brightness. The spatial distribution of the spectral index reveals that both non-thermal and thermal emission may be present in G5.97, dominating the non-thermal contribution in a major part of the source but  for the SW part where a gradient towards positive spectral index values is clearly seen. The thermal emission is too bright for a jet launched by the star associated to G5.97, considering the low mass loss rate expected for jets arising from proplyds. As the most likely explanation for the spectral index distribution, we propose that the shock created by the photo-evaporation flow with the collimated stellar wind of Her~36 accelerates charged particles producing non-thermal synchrotron emission in a considerable fraction of G5.97. According to this scenario, an electron density enhancement at the southwest of G5.97, maybe as a result of the orientation between the axis of the disk of the proplyd and Her~36, makes the thermal emission dominant over this region. 

\acknowledgments

JMM acknowledges financial support from DGAPA-UNAM through a postdoctoral fellowship. LFR acknowledges the support of DGAPA, UNAM, and of
CONACyT (M\'exico). 
The National Radio Astronomy Observatory is a facility of the National Science Foundation operated under cooperative agreement by Associated Universities, Inc.
We are grateful to Omaira Gonz\'alez-Mart\'in who reduced and
verified the X-ray data discussed in this paper.

\bibliography{paperG597}

\newpage

\input{obsG597.tex}
\input{fluxHer36.tex}
\input{flux_G597.tex}

\end{document}

%% file: obsG597.tex
\begin{table}[ht]
\scriptsize
\caption{Summary of the observations taken from the VLA archive.\label{observations}$^\mathrm{a}$}
\begin{center}
\begin{tabular}{lp{1cm}ccccccc}
\hline
\hline
 &                    &    & $\lambda$ & \multicolumn{2}{c}{Pointing center} & & Bootstraped Flux & Rms noise \\
 Date	 &  Conf. &  Project &    (cm)  &  	 $\alpha$~(J2000) & $\delta$~(J2000)	& Phase Calibrator     &     (Jy)  & ($10^{-4}$~Jy~beam$^{-1}$)   \\
\hline
1985 Apr  8   &  B &  AG178  &6.0 &$18^\mathrm{h}03^\mathrm{m}39\fs077$  & $-24\arcdeg23'10\farcs74$ & 1748-253 & $0.4773 \pm 0.0003$ &  4.5    \\ 
1986 Apr  27  &  A &  AW158  & 6.0 &$18^\mathrm{h}03^\mathrm{m}41\fs666$  & $-24\arcdeg22'40\farcs57$ & 1808-209& $0.3079 \pm 0.0004$     &   5.1   \\
1988 Jan 22 & B &  AT089  & 2.0 &$18^\mathrm{h}03^\mathrm{m}40\fs766$  & $-24\arcdeg22'40\farcs62$ & NRAO530 & $5.15 \pm 0.05$ &  4.2\\
1995 Sep 25  & BnA  &  AH557     & 6.0  &$18^\mathrm{h}03^\mathrm{m}41\fs565$  & $-24\arcdeg22'38\farcs56$ & 1730-130 &  $5.8 \pm 0.1$ &   5.0   \\ 
             & BnA  &  AH557     & 3.6 & $18^\mathrm{h}03^\mathrm{m}41\fs565$  & $-24\arcdeg22'38\farcs56$& 1730-130 &  $8.8 \pm 0.1$ &   5.7   \\ 
1996 Jan 23  & CnB  &  AH557     & 2.0 &$18^\mathrm{h}03^\mathrm{m}41\fs565$  & $-24\arcdeg22'38\farcs56$& 1730-130 &  $11.8 \pm 0.1$ &   5.5   \\
             & CnB  &  AH557      & 1.3 &$18^\mathrm{h}03^\mathrm{m}41\fs565$  & $-24\arcdeg22'38\farcs56$ & 1730-130 &  $14.36 \pm 0.02$ &  7.3   \\
1996 Dec 24 & A  &  AH605  &2.0  &$18^\mathrm{h}03^\mathrm{m}40\fs514$  & $-24\arcdeg22'44\farcs08$ & 1730-130& $17.04 \pm 0.13$ &  1.7\\
2003 Jul 5 &  A &  AB1094  & 6.0&$18^\mathrm{h}03^\mathrm{m}40\fs000$  & $-24\arcdeg22'40\farcs00$ & 1820-254 &  $0.7022 \pm 0.0006$  &  2.5\\
           &  A &   AB1094    & 3.6 &$18^\mathrm{h}03^\mathrm{m}40\fs000$  & $-24\arcdeg22'40\farcs00$ & 1820-254& $0.759 \pm 0.002$ &  1.2\\
2005 May 5 & B  &  AK586  & 1.3 &$18^\mathrm{h}03^\mathrm{m}40\fs000$  & $-24\arcdeg22'40\farcs00$  & 18210-25282 & $0.782 \pm 0.003$ &  3.2\\
2014 May 28 & A  &  14A-481  & 1.3 &$18^\mathrm{h}03^\mathrm{m}40\fs500$  & $-24\arcdeg22'44\farcs40$  & J1745-2900 & $0.992 \pm 0.001$ &  0.5\\ 
\hline
\end{tabular}
\end{center}
$^\mathrm{a}$The data are from the VLA archive, with the exception of the May 2014 data that was taken by us with the Jansky Very Large Array.
\end{table} 

%% file: fluxHer36.tex
\begin{table}[ht]
\footnotesize
\begin{center}
\caption{Parameters of the radio source associated with Her~36 SE\label{fluxHer36}.$^\mathrm{a}$}
\begin{tabular}{cccccc}
\hline
\hline
 $\alpha$~(J2000)$^\mathrm{b}$ & $\delta$~(J2000)$^\mathrm{b}$  &  Position Uncertainty$^\mathrm{c}$ &  $S_\mathrm{\nu}$ & &$\lambda$ \\
 ($18^h03^m$)  &  ($-24^\circ22'$)  & ($''$)  &  (mJy)      & Epoch  & (cm) \\
\hline
-- & -- & -- &  $\leq~1.35$ & 1985.3 & 6.0 cm \\ 
-- & -- & -- &  $\leq~1.53$ & 1986.3 & 6.0 cm \\ 
40.338$^s$ & 42\rlap{$''$}.86 & 0.14 &  $1.78 \pm 0.42$ & 1988.1 & 2.0 cm \\ 
-- & -- &-- &   $\leq~1.50$ & 1995.7 & 6.0 cm \\ 
-- & -- &-- &  $\leq~1.71$  & 1995.7 & 3.6 cm \\ 
-- & -- &-- &   $\leq~1.65$ & 1996.1 & 2.0 cm \\ 
-- & -- & -- &  $\leq~2.19$ & 1996.1 & 1.3 cm \\ 
40.338$^s$ & 42\rlap{$''$}.92 & 0.13& $1.79 \pm 0.40$ & 1997.0 & 2.0 cm$^\mathrm{d}$ \\ 
-- & -- & -- &  $\leq~0.75 $ & 2003.5 & 6.0 cm \\ 
40.340$^s$ & 42\rlap{$''$}.97 & 0.07 & $0.91 \pm 0.12 $ & 2003.5 & 3.6 cm \\ 
40.337$^s$ & 42\rlap{$''$}.99 & 0.13 &$1.45 \pm 0.32$ & 2005.3 & 1.3 cm \\
40.346$^s$ & 42\rlap{$''$}.83  &  0.02   &$1.22 \pm 0.05$ & 2014.4 & 1.3 cm$^\mathrm{e}$\\
\hline
\end{tabular}
\end{center}
$^\mathrm{a}$~Assuming that the source is not resolved.\\
$^\mathrm{b}$~Peak position measured on the map.\\
$^\mathrm{c}$~Obtained using the expression $\theta_s/(S_\mathrm{peak}/rms)$, where $\theta_s$ is the source size that, being unresolved, corresponds to the beamsize, and $S_\mathrm{peak}$ is the peak emission of Her 36 that is equivalent to $S_\mathrm{\nu}$, given in column 4. Finally, $rms$ is the root mean square
noise of the image.\\
$^\mathrm{d}$~We used the map restored with the beam of the January 1988 observations at 2.0~cm.\\ 
$^\mathrm{e}$~We used the map restored with the beam of the May 2005 observations at 1.3~cm.
\end{table}

%% file: flux_G597.tex
\begin{table}[ht]
\scriptsize
\begin{center}
\caption{Parameters of G5.97 resulting from a Gaussian fit (see text).$^\mathrm{a}$ \label{fluxG597}}
\begin{tabular}{p{1.4cm}p{1.4cm}cccccc}
\hline
\hline
   $\alpha$~(J2000) & $\delta$~(J2000)  &  Position Uncertainty &  $\theta_M  \times  \theta_m$ ; P.A.$^\mathrm{b}$  & $S_\mathrm{\nu}$ & &$\lambda$  \\
  ($18^h03^m$)  &  ($-24^\circ22'$)    & ({\bf$^s$},$''$)  &  ($'' \times ''$ ; deg.) &  (mJy)      & Epoch  & (cm) \\
\hline
40.517$^s$ & 44\rlap{$''$}.31 & (0.001 , 0.03) &  $ 0.75 \pm 0.28 \times 0.20 \pm 0.06; 135 \pm 133$ & $ 17.76 \pm 0.88$ & 1985.3 & 6.0 cm \\ 
40.510$^s$ & 44\rlap{$''$}.54 & (0.001 , 0.02) &  $ 0.46 \pm 0.06 \times 0.26 \pm 0.14; 137 \pm 20$ &  $ 14.11 \pm 0.85$ & 1986.3 & 6.0 cm \\ 
40.501$^s$ & 44\rlap{$''$}.29 & (0.001 , 0.02) &  $ 0.40 \pm 0.03 \times 0.30 \pm 0.05; 171 \pm 17$ &  $ 19.16 \pm 0.96$ & 1988.1 & 2.0 cm \\ 
40.515$^s$ & 44\rlap{$''$}.26 & (0.002 , 0.02) & unresolved or poorly determined & $ 15.80 \pm 0.88$ & 1995.7 & 6.0 cm \\ 
40.505$^s$ & 44\rlap{$''$}.24 & (0.002 , 0.02) & unresolved or poorly determined & $ 15.95 \pm 1.27$ & 1995.7 & 3.6 cm \\ 
40.498$^s$ & 44\rlap{$''$}.50 & (0.002 , 0.04) & unresolved or poorly determined &  $ 19.07 \pm 1.06$ & 1996.1 & 2.0 cm \\ 
40.486$^s$ & 44\rlap{$''$}.50 & (0.003 , 0.04) & unresolved or poorly determined & $ 17.89 \pm 1.98$ & 1996.1 & 1.3 cm \\ 
40.501$^s$ & 44\rlap{$''$}.40 & ($<$0.001 , 0.01) & $ 0.35 \pm 0.02 \times 0.23 \pm 0.04; 158 \pm 24$& $ 18.75 \pm 0.70$ & 1997.0 & 2.0 cm$^\mathrm{c}$ \\ 
40.505$^s$ & 44\rlap{$''$}.43 & (0.001 , 0.01) & $ 0.49 \pm 0.04 \times 0.25 \pm 0.11; 138 \pm 11$& $ 15.50 \pm 0.73$ & 2003.5 & 6.0 cm \\ 
40.503$^s$ & 44\rlap{$''$}.50 & ($<$0.001 , 0.01) & $ 0.29 \pm 0.01 \times 0.27 \pm 0.01; 53 \pm 29$ & $ 15.14 \pm 0.44$ & 2003.5 & 3.6 cm \\ 
40.501$^s$ & 44\rlap{$''$}.46 & (0.001 , 0.01) & $ 0.34 \pm 0.03 \times 0.25 \pm 0.05; 41 \pm 21$ & $ 15.12 \pm 0.59$ & 2005.3 & 1.3 cm \\ 
40.508$^s$ & 44\rlap{$''$}.34 & ($<$0.001 , 0.01) & $ 0.26 \pm 0.02  \times 0.23 \pm 0.02 ; 42 \pm 65$ & $ 12.56 \pm 0.24$ & 2014.4 & 1.3 cm$^\mathrm{d}$  \\ 
\hline
\end{tabular}
\end{center}
$^\mathrm{a}$~Errors are derived from the uncertainties given by the fitcomponents function of CASA.\\
$^\mathrm{b}$~The angular size is deconvolved with the synthesized beam of the corresponding observations.\\
$^\mathrm{c}$~We used the map restored with the beam of the January 1988 observations at 2.0~cm.\\ 
$^\mathrm{d}$~We used the map restored with the beam of the May 2005 observations at 1.3~cm.
\end{table}